\documentclass[12pt]{article}
\usepackage{graphics}
\usepackage{latexsym}
\usepackage[dvips,bookmarks]{hyperref}
\hypersetup{colorlinks,bookmarksopen,bookmarksnumbered,citecolor=blue,
	pdfstartview=FitH}
\def\hhref#1{\href{http://arxiv.org/abs/hep-th/#1}{hep-th/#1}}
\def\hpref#1{\href{http://arxiv.org/abs/hep-ph/#1}{hep-ph/#1}}
\def\haref#1{\href{http://arxiv.org/abs/astro-ph/#1}{astro-ph/#1}}
\def\hgref#1{\href{http://arxiv.org/abs/gr-qc/#1}{gr-qc/#1}}
\def\mhref#1{\href{mailto:#1}{#1}}
\begin{document}
\renewcommand{\theequation}{\thesection.\arabic{equation}}
\newcommand{\be}{\begin{equation}}
\newcommand{\ee}{\end{equation}}
\newcommand{\p}{\partial}
\newcommand{\hp}{\widehat{\p}}
\newcommand{\ov}{\overline}
\newcommand{\da}{^{\dagger}}
\newcommand{\w}{\wedge}
\newcommand{\st}{\stackrel}
\newcommand{\mb}{\mbox}
\newcommand{\mx}{\mbox}
\newcommand{\mt}{\mathtt}
\newcommand{\dt}{\mathtt{d}}
\newcommand{\al}{\alpha}
\newcommand{\bb}{\beta}
\newcommand{\ga}{\gamma}
\newcommand{\te}{\theta}
\newcommand{\Te}{\Theta}
\newcommand{\de}{\delta}
\newcommand{\et}{\tilde{e}}
\newcommand{\ze}{\xi}
\newcommand{\s}{\sigma}
\newcommand{\e}{\epsilon}
\newcommand{\om}{\omega}
\newcommand{\Om}{\Omega}
\newcommand{\la}{\lambda}
\newcommand{\La}{\Lambda}
\newcommand{\n}{\nabla}
\newcommand{\hn}{\widehat{\nabla}}
\newcommand{\hph}{\widehat{\phi}}
\newcommand{\ah}{\widehat{a}}
\newcommand{\bh}{\widehat{b}}
\newcommand{\ch}{\widehat{c}}
\newcommand{\ddh}{\widehat{d}}
\newcommand{\eh}{\widehat{e}}
\newcommand{\ph}{\widehat{p}}
\newcommand{\qh}{\widehat{q}}
\newcommand{\mh}{\widehat{m}}
\newcommand{\nh}{\widehat{n}}
\newcommand{\Dh}{\widehat{D}}
\newcommand{\stu}{\st{\sqcup}}
\newcommand{\au}{\stu{a}}
\newcommand{\bu}{\stu{b}}
\newcommand{\cu}{\stu{c}}
\newcommand{\du}{\stu{d}}
\newcommand{\eu}{\stu{e}}
\newcommand{\mmu}{\stu{m}}
\newcommand{\nnu}{\stu{n}}
\newcommand{\pu}{\stu{p}}
\newcommand{\Du}{\stu{D}}
\newcommand{\sto}{\st{\odot}}
\newcommand{\as}{\st{\odot}{a}}
\newcommand{\bs}{\st{\odot}{b}}
\newcommand{\cs}{\st{\odot}{c}}
\newcommand{\ds}{\st{\odot}{d}}
\newcommand{\es}{\st{\odot}{e}}
\newcommand{\ms}{\st{\odot}{m}}
\newcommand{\ns}{\st{\odot}{n}}
\newcommand{\ps}{\st{\odot}{p}}
\newcommand{\Ds}{\st{\odot}{D}}
\newcommand{\sts}{\st{\spadesuit}}
\newcommand{\sth}{\st{\heartsuit}}
\newcommand{\stp}{\st{\perp}}
\newcommand{\std}{\st{\diamondsuit}}
\newcommand{\ad}{\st{\spadesuit}{a}}
\newcommand{\bd}{\st{\spadesuit}{b}}
\newcommand{\cd}{\st{\spadesuit}{c}}
\newcommand{\gd}{\st{\spadesuit}{g}}
\newcommand{\dd}{\st{\spadesuit}{d}}
\newcommand{\Dd}{\st{\spadesuit}{D}}
\newcommand{\ed}{\st{\spadesuit}{e}}
\newcommand{\fd}{\st{\spadesuit}{f}}
\newcommand{\zd}{\st{\spadesuit}{\xi}}
\newcommand{\md}{\st{\spadesuit}{m}}
\newcommand{\nd}{\st{\spadesuit}{n}}
\newcommand{\stc}{\st{\clubsuit}}
\newcommand{\az}{\st{\clubsuit}{a}}
\newcommand{\bz}{\st{\clubsuit}{b}}
\newcommand{\cz}{\st{\clubsuit}{c}}
\newcommand{\dz}{\st{\clubsuit}{d}}
\newcommand{\Dz}{\st{\clubsuit}{D}}
\newcommand{\ez}{\st{\clubsuit}{e}}
\newcommand{\fz}{\st{\clubsuit}{f}}
\newcommand{\nz}{\st{\clubsuit}{n}}
\newcommand{\mz}{\st{\clubsuit}{m}}
\newcommand{\tb}{\overline{\theta}}
\newcommand{\ti}{\widetilde}

\newcommand{\2}{\textstyle{1\over 2}}
\newcommand{\3}{\frac{1}{3}}
\newcommand{\4}{\frac{1}{4}}
\newcommand{\8}{\frac{1}{8}}
\newcommand{\6}{\frac{1}{16}}

\newcommand{\ra}{\rightarrow}
\newcommand{\Ra}{\Rightarrow}
\newcommand{\im}{\Longleftrightarrow}
\newcommand{\hs}{\hspace{5mm}}
\newcommand{\x}{\star}
\newcommand{\Delt}{\p^{\star}}

\thispagestyle{empty}
{\bf 21st May, 2003} \hspace{\fill}
{\bf YITP-SB-03-18}

\vspace{1cm}
\begin{center}{\Large{\bf QUINTESSENTIAL TRANSITION OF THE \\

\vspace{3mm}
 INTERNAL MANIFOLD}}\\
\vspace{1cm}
{\large{\bf T. Biswas\footnote{\mhref{tirtho@insti.physics.sunysb.edu} }}}\\
\vspace{5mm}
{\small C.N. Yang Institute of Theoretical Physics\\
Department of Physics and Astronomy\\
State University of New York at Stony Brook\\
Stony Brook, New York 11794-3840}
\end{center}

\begin{abstract}
In this paper, I study the dynamics of the Kaluza-Klein internal  manifold using its ``shape'' and the ``size'' as collective coordinates. The essential motivation is to be able to explain symmetry breaking in gauge theories through a transition of the internal manifold  from a symmetrical metric space to a less symmetric one. We find that this may be possible depending on the values of certain group theoretical parameters. Further, this transition resembles a ``quintessential inflation'' scenario, the feasibility of which however, needs to be further studied in details.  
\end{abstract}

\newpage
\setcounter{page}{1}

\section{{\bf   INTRODUCTION}}

Kaluza-Klein theories provide an elegant way of combining (four-dimensional) gravity with gauge interactions in a  geometric way through dimensional reduction  schemes (see for example \cite{kaluza} for details). One starts with  a higher dimensional space-time containing four dimensional observable or ``external'' universe along with  extra dimensions constituting  the ``internal manifold''. The latter remains unobserved essentially due to its smallness\footnote{In the brane world scenario \cite{braneworld} large extra dimensions are also possible.}. One then usually considers the vacuum to be a product of a four-dimensional vacuum manifold (Minkowski, deSitter or anti deSitter), and a  compact internal manifold with matching scalar curvature constants. Four-dimensional physics then arises as fluctuations around this vacuum. For example, if one looks at the massless modes, which are  important for describing low energy physics of the higher dimensional metric, then one finds a  graviton (in the four-dimensional sector of the metric) and  gauge bosons (appearing in the off-diagonal part of the metric) associated with the Killing vectors of the ``frozen'' internal manifold. The symmetries of the internal manifold translates into gauge symmetry in the observed four dimensional universe. It is natural then to suspect that when we observe a symmetry breaking in nature (Standard Model for example), we are really observing a shadow of a symmetry breaking taking place in the internal manifold;  a dynamical transition  from a more symmetric internal space (``spherical'') to a less symmetric (``squashed'') one. In this paper I show that this indeed may be the case where the internal manifold starts off with a symmetric metric, rolls over a potential barrier (or tunnels through) to reach a  squashed state! Further, the dynamics  resembles the now well known  scenario of ``quintessential inflation'' \cite{peebles}, where the rolling over phase corresponds to a deSitter like inflation, and  at late times after the transition, the internal manifold keeps getting more and more squashed accompanied by quintessence. This picture departs fundamentally from the concept of a frozen internal manifold to that of a  dynamic one. 

Previously geometric mechanisms of symmetry breaking have been realized by introducing additional scalar fields \cite{sobczyk}. However, I concentrate only on  pure Kaluza-Klein gravity (no extra non-geometric scalar fields) where the initial isometry group $G_L\otimes G_R$ of the internal manifold (which is a Lie group say $G$) is broken down to  $G_L\otimes H_R$ \cite{t}. This should perhaps be viewed only as a toy model to be extended to supergravity (SUGRA). It should be mentioned that  ideas of using geometry of extra dimensions to break gauge symmetry can also be found  in the context of ``dimensional reduction by isometries'' \cite{cho}, which is however fundamentally different from the Kaluza-Klein scenario. 

Within the Kaluza-Klein framework geometric ways of breaking symmetry  has also been studied for some special internal manifolds using a quantum field theoretic approach \cite{okada}. However, in this paper I study the dynamics from a cosmological view point. As was suggested in \cite{t} I first identify the scalar fields corresponding to the size and the shape of the internal manifold that are relevant to study the phase transition. I obtain an effective action of these scalar fields coupled to four dimensional gravity and show that the truncation is consistent \cite{duff}, i.e. the solutions of the field equations derived from the effective action are also solutions of the complete higher dimensional Einstein's equations. One can then derive a quantum mechanical action by treating these  fields  as collective coordinates characterizing the internal manifold.

To study the dynamics comprehensively is a difficult and arduous task but one can get significant insight by looking at the field equations, effective potentials and approximate solutions. The effective quantum mechanical action contains three variables, the radii of our observational and internal dimensions, $A(t)$ and $S(t)$ respectively, and a squashing variable, $T(t)$. We are specially interested in understanding the behaviour of $T(t)$ which plays a pivotal role in the phase transition. It turns out that one can define an effective potential for $T(t)$ which provides a  qualitative picture of its dynamics. Depending on the group theoretical parameters there are two possible cases: Say $\stc{D}$ and $\sts{D}$ label the dimensions of the coset space $G/H$ and the subgroup $H$ respectively. Then for $\stc{D}\ >2\sts{D}$ one  finds only a single stable minima  corresponding to the usual symmetric  Kaluza-Klein vacuum. No geometric transition is obviously possible. However for  $\stc{D}\ <2\sts{D}$ there is a minima (symmetric vacuum) and  a maxima (the squashed Einstein space) after which the potential falls  and then rolls toward zero asymptotically. This case  may realize the scenario of quintessential inflation. 

The inflationary mechanism was first developed \cite{guth} to grapple with some of the long standing problems of standard cosmology like the flatness and horizon problems and has now been found to possess several other virtues (see for example \cite{linde} for a review). In essence, the inflationary model posits a phase of exponential growth in our early universe which makes it possible to be much flatter, larger, and more causally connected. In the model we are investigating,  the internal manifold starts out in the  symmetric  vacuum state. It  then rolls over the squashed vacuum $dS_4\otimes G_{\mt{sq}}$. This signals an inflationary phase, the elevated potential energy of the scalar fields acting as an effective cosmological constant. Once the inflation ends (i.e. the internal manifold makes the transition to the squashed phase) presumably the universe enters a radiation-matter dominated era. However, one quickly realises that  the later phase of scalar evolution rolling down the potential toward zero can give rise to quintessence. Generically speaking quintessence is the phenomena \cite{ratra} when the scalar field(s) slowly rolls down a potential hill toward zero so that we observe a small effective cosmological constant originating in the small potential energy of the scalar field(s). This was basically invented to solve the fine tuning problem of the cosmological constant since its value is so small as compared to the typical mass scales in the early universe. It has recently become popular in the light of the experimental observation \cite{exp} that the universe may be accelerating with a small positive cosmological constant; in other words, a near exponential growth which can be explained by quintessence. In our case $T(t)$ rolls along an exponential potential \cite{liddle,copeland} thus making quintessence a viable prospect. Indeed, we find such an approximate solution of late time exponential growth of our observed universe. 

Could it be that the  geometric symmetry breaking mechanism in Kaluza-Klein theories can also explain inflation and quintessence?

To answer this question more comprehensively one has to firstly, put in matter-radiation contributions to Einstein's equations and carry out  a much more rigorous and probably numerical analysis addressing issues like primordial density fluctuations, baryogenesis \cite{joyce}, nucleosynthesis \cite{liddle}, relic particle abundance \cite{relic}, gravitational waves \cite{peebles} etc, which is however  out of the scope of this paper. Also, Kaluza-Klein reduction schemes in general suffers from the problem of expanding size, $S(t)$, of the internal manifold. Clearly for this whole scheme to be consistent one needs a confining force for the internal manifold. In the context of string theory it was realized that strings wrapping around compact circles tend to compress it much like an elastic band \cite{brane}. This phenomena was later generalized to include higher dimensional branes \cite{easson}.  Recent progress in brane-cosmology suggests that pressureless brane gas may be the key to understanding the puzzle of stabilizing the size of the internal manifold. It is therefore important to incorporate this effect in our analysis as $S(t)$ and $T(t)$ evolutionary equations are intricately interlinked. In this context it is worth mentioning that recently attempts have been made to realize an accelerating universe involving the dynamics of the size(s) \cite{size} of the extra dimensions. It should be contrasted with the shape that, I propose here, plays the key role in the cosmological dynamics and phase transition.

The paper is organised as follows: In section 1, I identify the relevant scalar fields  in the higher dimensional metric and obtain an effective action through dimensional reduction. In section 2, I first check consistency of this action and then proceed to obtain a quantum mechanical action involving the shape and size of our universe. In the next section I choose a specific simplifying gauge and obtain field equations. In section 4, I study the dynamics and its implications toward inflation and quintessence. Finally, I conclude with a brief summary and some remarks about future research directions.   

\section{{\bf  OUR MODEL }}
As in \cite{t} we consider our universe to be a semi-direct product, $M_{D+1}\otimes G$, where $M_{D+1}$ is the $D+1$-dimensional observational universe and $G$, a Lie group manifold, serves as the Kaluza-Klein internal space \cite{kerner}. Before we decide on an ``ansatz'' for the dimensional reduction, let us quickly review the Lie group geometry. 
\vspace{5mm}
\\
{\bf Geometry of Lie groups:}  A Lie group element $g$ can be parameterised as 
\begin{equation}
g=exp(\chi^{\as}(y^{\ms})T_{\as})\ \in\ G
\end{equation}
where $T_{\as}\ \in\ \cal{G}$, the Lie algebra corresponding to the Lie group $G$ and $\chi^{\as}(y^{\ms})$ are some given functions of the coordinates $y^{\ms}$ charting the Lie group manifold. The Lie group generators $T_{\as}$ satisfy the usual commutation relations:
\begin{equation}
[T_{\as},T_{\bs}]=C_{\as\bs}{}^{\cs}T_{\cs}
\end{equation}
where $C_{\as\bs}{}^{\cs}$ are the structure constants of the Lie group. Further, since we are principally interested in a transition of the group manifold from a $G_L\otimes G_R$ maximally symmetric metric to a $G_L\otimes H_R$ invariant metric, we suitably choose our generators $\{T_{\as}\}=\{T_{\az},T_{\ad}\}$ such that $\{T_{\ad}\}$ span $\cal{H}$. We will always refer group quantities by a circle ($\odot$) while that of the  {\bf{\em C}}oset space $G/H$ and the {\bf {\em S}}ubgroup $H$ with {\bf {\em C}}lubs ($\clubsuit$) and {\bf {\em S}}pades ($\spadesuit$) respectively. Sometimes we may omit the symbols when it is self-evident. We will also assume the groups $G$ and $H$ to be simple and the coset decomposition to be reductive and symmetric
\be
C_{\az\bd}{}^{\cd}=C_{\az\bz}{}^{\cz}=0
\ee
If we choose the left invariant vector fields $\{e_{\as}\}$ as vielbein basis vectors, a generic  $G_L\otimes H_R$-invariant metric is a constant (in the flat basis) satisfying (see for example \cite{t} for details)
\be
D_{\as}{}^{\cs}(h)D_{\bs}{}^{\ds}(h)g_{\cs\ds}=g_{\as\bs};\ \forall h\in H
\ee
$D_{\as}{}^{\bs}(g)$ being the adjoint representation of $G$. In our case such a metric looks like
\begin{equation}
g^S_{\as\bs}=\left( \begin{array}{cc}
g^K_{\az\bz} & 0\\
0& T^2g^K_{\ad\bd}
\end{array} \right)
\end{equation}
where $T^2$ is a constant ``squashing'' parameter and $g_K$ is the Killing metric. Clearly, for $T=1$ we recover the Killing metric which turns out to be $G_L\otimes G_R$ invariant. The Killing metric is consistent with its usual identification as the Kaluza-Klein vacuum metric as it is  Einstein, i.e. it satisfies
\begin{equation}
R_{\as\bs}=\sto{\la}g_{\as\bs}
\label{eq:4einstein}
\end{equation}
The constant $\sto{\la}$ is often known as the internal curvature. Contrary to this picture of an  internal manifold frozen in its maximally symmetric Killing metric, we treat it as  dynamic. In particular, we want to study whether the manifold makes a transition from $T=1$ (symmetric case) to $T\neq 1$ metric. 

For some values of the squashing parameter, other than 1, we can also have a Einstein manifold. Thus in \cite{t} it was suggested that the internal manifold may make a transition from say the maximally symmetric Einstein space to the less symmetric one. It is clear, what we have to do to understand this dynamics; we should treat $T$ as a collective coordinate $T(t)$ charecterizing the shape of the group manifold. We know to have a consistent dimensional reduction ansatz one has to also include the overall size $S(t)$ of the internal manifold. Thus our ansatz for the group metric will be given by \cite{t} 
\begin{equation}
g_{\as\bs}(t)=S^2(t)\left( \begin{array}{cc}
g^K_{\az\bz} & 0\\
0& T^2(t)g^K_{\ad\bd}
\end{array} \right)
\end{equation}
{\bf Field Theoretic Ansatz:} In the Kaluza-Klein reduction scheme we now know exactly which scalars are relevant to studying the dynamics of symmetry breaking, viz. $S(t)\ra \Psi(x)$ and $T(t)\ra \Te(x)$. We will denote the coordinates charting the observable universe $M_{D+1}$ by $x^m$ while $x^{\mh}$ will be used to collectively  denote $\{x^m,y^{\ms}\}$. We will also use ``hatted'', $\widehat{ }$,  quantities to refer to objects corresponding to the full higher dimensional manifold. 

Although an expression of the metric of the form (1.7) is physically clarifying, technically it is more convenient to include the scalars in the vielbein. We choose to  parameterise the group element as 
$$g=exp(\chi^{\az}(y^{\mz})T_{\az}) exp(\chi^{\ad}(y^{\md})T_{\ad})$$ The ansatz for the full higher dimensional vielbein is then given by
\begin{equation}
\eh_{\mh}{}^{\ah}=\left( \begin{array}{ccc}
e_m{}^{a}(x) & 0&0\\
0 &\Psi(x)\es_{\mz}{}^{\az}(y)&\Psi(x)\Te(x)\es_{\mz}{}^{\ad}(y)\\
0&0& \Psi(x)\Te(x)\es_{\md}{}^{\ad}(y)
\end{array} \right)
\label{eq:4s-inv}
\ee
and
\be
\eh_{\ah}{}^{\mh}=\left( \begin{array}{ccc}
e_{a}{}^{m}(x) & 0&0\\
0 &\ \Psi^{-1}(x)\es_{\az}{}^{\mz}(y)&\Psi^{-1}(x)\es_{\az}{}^{\md}(y)\\
0&0& \Psi^{-1}(x)\Te^{-1}(x)\es_{\ad}{}^{\md}(y)
\end{array} \right)
\label{eq:4s-vielbein}
\end{equation}
The flat-metric is then just a constant
\begin{equation}
\widehat{g}_{\ah\bh}=\left( \begin{array}{cc}
g_{ab} & 0\\
0 & g^K_{\as\bs}
\end{array} \right)
\label{eq:4s-metric}
\end{equation}
We did not include the vectors in the ansatz (1.8-1.9) because we are only interested in the vacuum dynamics and the vectors appear as fluctuations around the vacuum metric. The vacuum dynamics is of course relevant to the gauge vectors as, if the transition occurs, the gauge symmetry will be broken from $G_R\ra H_R$, with the broken gauge bosons acquiring mass as explained in \cite{t}.\\ 
\vspace{5mm}
\\ 
{\bf Effective Action:} Our next task is to obtain an effective action for the ansatz (\ref{eq:4s-inv}-\ref{eq:4s-metric}) via dimensional reduction of the higher dimensional gravitational action\be
\widehat{S}_{\widehat{D}}=\int dx^{\widehat{D}}\ \eh^{-1}\widehat{R}
\label{eq:4action}
\ee
In order to compute the scalar curvature $\widehat{R}$ we first need to compute the spin connections $\widehat{\om}^{\ah}{}_{\bh}$ which are uniquely defined via
\be
\ddh \widehat{\om}^{\ah}+\widehat{\om}^{\ah}{}_{\bh}\w\widehat{\om}^{\bh}=0
\label{eq:4connection}
\ee
where $\widehat{\om}^{\ah}$ are the basis 1-forms
\be
\widehat{\om}^{\ah}=dx^{\mh}e_{\mh}{}^{\ah}
\ee
For (\ref{eq:4s-inv}) the 1-forms are given by
$$\widehat{\om}^{a}=\om^a$$
$$\widehat{\om}^{\az}=\Psi\om^{\az}$$
\be
\widehat{\om}^{\ad}=\Psi\Te\om^{\ad}
\ee
With a little algebraic manipulation and guess work one can obtain the connections satisfying (\ref{eq:4connection}):
$$\widehat{\om}^{a}{}_{b}=\om^{a}{}_{b}$$
$$\widehat{\om}^{\az}{}_{b}=(e_b\Psi)\om^{\az}$$
$$\widehat{\om}^{\ad}{}_{b}=(e_b\Psi\Te)\om^{\ad}$$
$$\widehat{\om}^{\az}{}_{\bz}=\om^{\az}{}_{\bz}+(\Te^2-1)\om^{\az}{}_{\cd\bz}\om^{\cd}$$
$$\widehat{\om}^{\az}{}_{\bd}=\Te\om^{\az}{}_{\bd}$$
\be
\widehat{\om}^{\ad}{}_{\bd}=\om^{\ad}{}_{\bd}
\ee
$e_b$ is the vielbein vector
$$e_b=e_b{}^m\p_m$$
and $\om^{\as}{}_{\bs\cs}$'s are the group connection co-efficients defined by
\be
\om^{\as}{}_{\bs\cs}=g^{\as\as'}\om_{\as'\bs\cs}\ ;\ \om_{\as\bs\cs} =\2(C_{\as\bs\cs}+C_{[\as\cs\bs]})
\ee
Our next step is to evaluate the curvature 2-forms
\be
\widehat{\cal{R}}^{\ah}{}_{\bh}=\ddh \widehat{\om}^{\ah}{}_{\bh}+\widehat{\om}^{\ah}{}_{\ch}\w\widehat{\om}^{\ch}{}_{\bh}
\ee
A straight forward computation yields the following results:
$$\widehat{\cal{R}}^{a}{}_{b}=\mathcal{R}^{a}{}_{b}$$
$$\widehat{\cal{R}}^{\az}{}_{b}=\Psi^{-1}\n_c(e_b\Psi)\widehat{\om}^{c}\w\widehat{\om}^{\az}+\Psi^{-1}(e_b\Te)\om^{\az}{}_{\cd\dz}\widehat{\om}^{\dz}\w\widehat{\om}^{\cd}$$
$$\widehat{\cal{R}}^{\az}{}_{\bz}=\mathcal{R}^{\az}{}_{\bz}$$
$$+\{(\Te^2-1)[\om^{\az}{}_{\cd\dz}\om^{\dz}{}_{\bz\ed}-\om^{\dz}{}_{\cd\bz}\om^{\az}{}_{\dz\ed}+\om^{\az}{}_{\fd\bz}\om^{\fd}{}_{\cd\ed}]+(\Te^2-1)^2\om^{\az}{}_{\cd\dz}\om^{\dz}{}_{\ed\bz})\}\Psi^{-2}\Te^{-2}\widehat{\om}^{\cd}\w\widehat{\om}^{\ed}$$
$$+\{(\Te^2-1)(\om^{\az}{}_{\cd\dz}\om^{\dz}{}_{\cd\ez}+\om^{\az}{}_{\cd\bz}\om^{\cd}{}_{\dz\ez})-(\p\Psi)^2\de_{\dz}{}^{\az}g_{\ez\bz}\}\Psi^{-2}\widehat{\om}^{\dz}\w\widehat{\om}^{\ez}+2\Psi^{-2}(e_d\Te)\om^{\az}{}_{\cd\bz}\widehat{\om}^{d}\w\widehat{\om}^{\cd}$$
$$\widehat{\cal{R}}^{\ad}{}_{b}=\Psi^{-1}\Te^{-1}\n_c(e_b\Psi\Te)\widehat{\om}^{c}\w\widehat{\om}^{\ad}+\Psi^{-1}(e_b\Te)\om^{\ad}{}_{\cz\dz}\widehat{\om}^{\cz}\w\widehat{\om}^{\dz}$$
$$\widehat{\cal{R}}^{\ad}{}_{\bd}=\mathcal{R}^{\ad}{}_{\bd}-2\Psi^{-2}\Te^{-2}(\p\Te\Psi)^2\widehat{\om}^{\ad}\w\widehat{\om}_{\bd}+\Psi^{-2}(\Te^2-1)\om^{\ad}{}_{\cz\dz}\om^{\cz}{}_{\bd\ez}\widehat{\om}^{\dz}\w\widehat{\om}^{\ez}$$
$$
\widehat{\cal{R}}^{\az}{}_{\bd}=\Te\mathcal{R}^{\az}{}_{\bd}+\Psi^{-1}e_c\Te\om^{\az}{}_{\bd\dz}\widehat{\om}^{c}\w\widehat{\om}^{\dz}-\Psi^{-2}\Te^{-1}(e_c\Psi)(e^c\Te\Psi)\widehat{\om}^{\az}\w\widehat{\om}^{\bd}$$
\be
+\Psi^{-2}(\Te^2-1)\om^{\az}{}_{\dd\cz}\om^{\cz}{}_{\bd\ez}\widehat{\om}^{\dd}\w\widehat{\om}^{\ez}
\ee
The coefficients of the Riemannian tensor can now be read off  from the curvature 2-forms
\be
\widehat{\cal{R}}^{\ah}{}_{\bh}=\widehat{R}^{\ah}{}_{\bh|\ch\ddh|}\widehat{\om}^{\ch}\w\widehat{\om}^{\ddh}
\ee
Here $|\ |$ indicates that the sum counts a pair only once. The Riemannian coefficients obtained thus are
$$\widehat{R}^{a}{}_{bcd}=R^{a}{}_{bcd}$$
$$\widehat{R}^{\az}{}_{bc\dz}=\Psi^{-1}\n_c(e_b\Psi)\de_{\dz}{}^{\az}$$
$$\widehat{R}^{\az}{}_{b\cz\dd}=\Psi^{-1}e_b\Te\om^{\az}{}_{\dd\cz}$$
$$\widehat{R}^{\az}{}_{\bz\cz\dz}=\Psi^{-2}[R^{\az}{}_{\bz\cz\dz}+(\Te^2-1)(\om^{\az}{}_{\ed[\cz}\om^{\ed}{}_{\bz\dz]}+\om^{\az}{}_{\ed\bz}\om^{\ed}{}_{[\cz\dz]})-(\p\Psi)^2\de_{[\cz}{}^{\az}g_{\dz]\bz}]$$
$$\widehat{R}^{\az}{}_{\bz\cz\dd}=\Psi^{-2}\Te^{-1}R^{\az}{}_{\bz\cz\dd}$$
$$\widehat{R}^{\az}{}_{\bz\cd\dd}=\Psi^{-2}\Te^{-2}[R^{\az}{}_{\bz\cd\dd}+(\Te^2-1)(\om^{\az}{}_{[\cd\ez}\om^{\ez}{}_{\bz\dd]}-\om^{\ez}{}_{[\cd\bz}\om^{\az}{}_{\ez\dd]}+\om^{\az}{}_{\ed\bz}\om^{\ed}{}_{[\cd\dd]})$$
$$+(\Te^2-1)^2\om^{\az}{}_{[\cd\ez}\om^{\ez}{}_{\dd]\bz}]$$
$$\widehat{R}^{\az}{}_{\bz c\dd}=2\Psi^{-1}(e_c\Psi)\om^{\az}{}_{\dd\bz}$$
$$\widehat{R}^{\ad}{}_{bc\dd}=\Psi^{-1}\Te^{-1}\n_c(e_b\Psi\te)\de_{\dd}{}^{\ad}$$
$$\widehat{R}^{\ad}{}_{b \cz\dz}=\Psi^{-1}(e_b\Te)\om^{\ad}{}_{[\cz\dz]}$$
$$\widehat{R}^{\ad}{}_{\bd\cz\dz}=\Psi^{-2}[R^{\ad}{}_{\bd\cz\dz}+(\Te^2-1)\om^{\az}{}_{\ez[\cz}\om^{\ez}{}_{\bd\dz]}]$$
$$\widehat{R}^{\ad}{}_{\bd\cd\dz}=\Psi^{-2}\Te^{-1}R^{\ad}{}_{\bd\cd\dz}$$
$$\widehat{R}^{\ad}{}_{\bd\cd\dd}=\Psi^{-2}\Te^{-2}[R^{\ad}{}_{\bd\cd\dd}-(\p\Psi\Te)^2\de_{[\cd}{}^{\ad}g_{\bd\dd]}]$$
$$\widehat{R}^{\az}{}_{\bd\cz\dz}=\Psi^{-2}\Te R^{\az}{}_{\bd\cz\dz}$$
$$\widehat{R}^{\az}{}_{\bd\cd\dz}=\Psi^{-2}[R^{\az}{}_{\bd\cd\dz}+(\Te^2-1)\om^{\az}{}_{\cd\ez}\om^{\ez}{}_{\bd\dz}+\Psi^{-1}(\p\Psi)(\p\Psi\Te)\de_{\dz}{}^{\az}g_{\bd\cd}]$$
$$\widehat{R}^{\az}{}_{\bd\cd\dd}=\Psi^{-2}\Te^{-1}R^{\az}{}_{\bd\cd\dd}$$
\be
\widehat{R}^{\az}{}_{\bd c\dz}=\Psi^{-1}(e_c\Te)\om^{\az}{}_{\bd\dz}
\ee
From the Riemann tensor it is easy to calculate the Ricci tensor
\be
\widehat{R}_{\bh \ddh}=\widehat{R}^{\ah}{}_{\bh \ah\ddh}
\ee
After some simplifications one obtains
$$\widehat{R}_{a b}=R_{a b}-\Ds\Psi^{-1}\n_b(e_a\Psi)-\Dd\Te^{-1}\n_b(e_a\Te)-\Dd\Psi^{-1}\Te^{-1}e_{(a}\Psi e_{b)}\Te$$
$$\widehat{R}_{\az \bz}=g_{\az \bz}[-\{\Psi^{-1}\Box\Psi+(\Ds-1)\Psi^{-2}(\p\Psi)^2+\Dd\Psi^{-1}\Te^{-1}\p_{a}\Psi \p^{a}\Te\}+\Psi^{-2}\sto{\la}-\4\Psi^{-2}(\Te^2-1)]$$
$$
\widehat{R}_{\ad \bd}=g_{\ad \bd}[-\{\Psi^{-1}\Box\Psi+\Te^{-1}\Box\Te+(\Ds-1)\Psi^{-2}(\p\Psi)^2+(\Dd-1)\Te^{-2}(\p\Te)^2$$
\be
+(\Dd+\Ds)\Psi^{-1}\Te^{-1}\p_{a}\Psi \p^{a}\Te\}+\Psi^{-2}(\sto{\la}-k\sts{\la})+\4\Psi^{-2}(\Te^2-1)(1-k)+\Psi^{-2}\Te^{-2}k\sts{\la}]
\label{eq:4ricci}
\ee
Here we have introduced a group theoretical parameter k:
\be
\stc{g}_{\ad \bd}=k\sts{g}_{\ad \bd}
\ee
$\sts{g}_{\ad \bd}$ is the Killing metric of group $H$ while $\stc{g}_{\ad \bd}$  of course corresponds to the Killing metric of group $G$. For a symmetric coset decomposition it is known that
\be
k=1-\frac{\Dz}{2\Dd}
\ee
Also, $\sts{\la}$ is defined in the usual way as in (\ref{eq:4einstein}) except that now all the quantities refer to the subgroup $H$. In fact for Killing metrics 
$$\sts{\la}=\sto{\la}=\4$$
and we will explicitly substitute their values.

We are ready to compute the scalar curvature that we need in the action.
\be
\widehat{R}=g^{\ah \bh}\widehat{R}_{\ah \bh}=g^{a b}\widehat{R}_{a b}+g^{\az \bz}\widehat{R}_{\az \bz}+g^{\ad \bd}\widehat{R}_{\ad \bd}
\ee
Finally, we have
$$\widehat{R}=R-\left[2\Ds\frac{\Box\Psi}{\Psi}+2\Dd\frac{\Box\Te}{\Te}+\Ds(\Ds-1)\frac{(\p\Psi)^2}{\Psi^2}+\Dd(\Dd-1)\frac{(\p\Te)^2}{\Te^{2}}+2\Dd(\Ds+1)\frac{\p_{a}\Psi \p^{a}\Te}{\Psi\Te}\right]$$
\be
+\4\left[  2\Dz\frac{1}{\Psi^2}-\Dz \frac{\Te^2}{2\Psi^2}+ k\Dd \frac{1}{\Psi^2\Te^2} \right]
\label{eq:4curvature}
\ee
Since $\widehat{R}$ is independent of the group coordinates one can perform the integration over the group in the action (\ref{eq:4action}) which essentially just yields a volume factor $V_G$. Thus we have our effective $D+1$-dimensional action
\be
S_{\mt{grav}}=V_G\int e^{-1}\Psi^{\stackrel{\odot}{D}}\Te^{\Dd}\widehat{R}
\ee
It is useful to perform some integration by parts. The simplified action looks like
\be
S_{\mt{grav}}=V_G\int dx^{D+1}\ e^{-1}\Psi^{\Ds}\Te^{\Dd}[R-K+V]
\ee
where we have defined the Kinetic and Potential like terms for the scalar fields as
\be
K=-\left[\Ds(\Ds-1)\frac{(\p\Psi)^2}{\Psi^2}+\Dd(\Dd-1)\frac{(\p\Te)^2}{\Te^{2}}+2\Dd(\Ds-1)\frac{\p_{a}\Psi \p^{a}\Te}{\Psi\Te}\right]
\ee
and
\be
V=\4\left[ 2\Dz\frac{1}{\Psi^{2}}-\Dz\frac{\Te^2}{2\Psi^{2}}+k\Dd\frac{1}{\Psi^{2}\Te^{2}}\right]
\ee
At this point it is useful to redefine the scalars:
\be
\Psi=e^{\psi};\mbox{ and }\Te=e^{\te}
\ee
The kinetic and potential terms then look like
\be
K=-\left[\Ds(\Ds-1)(\p\psi)^2+\Dd(\Dd-1)(\p\te)^2+2\Dd(\Ds-1)\p_{a}\psi \p^{a}\te\right]
\ee
and
\be
V=\4\left[ 2\Dz e^{-2\psi}-\2\Dz e^{2(\te-\psi)}+k\Dd e^{-2(\psi+\te)}\right]
\ee
The action is given by
\be
S_{\mt{grav}}=V_G\int dx^{D+1}\ e^{-1}e^{\Ds\psi+\Dd\te}[R-K+V]
\label{eq:4e-action}
\ee
Finally, one can also include a cosmological term in the higher dimensional action
\be
\widehat{S}_{\mt{cos}}=-2\La\int dx^{\widehat{D}}\ \eh^{-1}
\ee
The corresponding term in the effective action is
\be
S_{\mt{cos}}=-2\La V_G\int dx^{D+1}\ e^{-1}e^{\Ds\psi+\Dd\te}
\label{eq:4c-action}
\ee
\setcounter{equation}{0}
\section{ {\bf CONSISTENCY AND THE QUANTUM MECHANICAL EFFECTIVE ACTION }}
In the previous section we obtained the dimensionally reduced field theoretic action for our model. It is important that we check the consistency of our ansatz \cite{duff}. We should check that the solutions that we obtain by varying the effective action (1.35) are indeed solutions of the full higher dimensional Einstein's equations, and this would mean that the truncation we performed is legitimate.
\vspace{5mm}
\\ 
{\bf Consistency of the Truncation:} To obtain Einstein's field equations we essentially have to compute the Einstein tensor
\be
\widehat{G}_{\ah\bh}=\widehat{R}_{\ah\bh}-\2\widehat{R}\widehat{g}_{\ah\bh}
\ee
Using (\ref{eq:4ricci}) and (\ref{eq:4curvature}) we obtain
\be
\widehat{G}_{ab}=\widehat{R}_{ab}-\2 g_{ab}\widehat{R}
\label{eq:4Gab}
\ee
$$
\widehat{G}_{\az\bz}=\sto{g}_{\az\bz}\left[(\Ds-1)\Psi^{-1}\Box\Psi+\Dd\Te^{-1}\Box\Te+\2(\Ds-1)(\Ds-2)\Psi^{-2}(\p\Psi)^2 \right.$$
$$+\2\Dd(\Dd-1)\Te^{-2}(\p\Te)^2+\Dd\Ds\Psi^{-1}\Te^{-1}\p_{a}\Psi \p^{a}\Te$$
\be
\left.+\Psi^{-2}\4(2-\Dz)-\8\Dd k\Psi^{-2}\Te^{-2}+\6(\Dz-4)\Psi^{-2}\Te^{2}-\2 R\right]
\label{eq:4Gabc}
\ee
and
$$
\widehat{G}_{\ad\bd}=\sto{g}_{\ad\bd}\left[(\Ds-1)\Psi^{-1}\Box\Psi+(\Dd-1)\Te^{-1}\Box\Te+\2(\Ds-1)(\Ds-2)\Psi^{-2}(\p\Psi)^2 \right.$$
$$+\2(\Dd-1)(\Dd-2)\Te^{-2}(\p\Te)^2+(\Dd-1)\Ds\Psi^{-1}\Te^{-1}\p_{a}\Psi \p^{a}\Te$$
\be
\left.-\4\Dz\Psi^{-2}-\8(2-\Dd) k\Psi^{-2}\Te^{-2}+\6\frac{\Dz}{\Dd}(\Dd+2)\Psi^{-2}\Te^{2}-\2 R\right]
\label{eq:4Gabs}
\ee
The pure gravity field equations read
\be
 \widehat{G}_{\ah\bh}=0
\ee
Our task is to show that the field equations that one obtains by varying the effective action (\ref{eq:4e-action}) also satisfies (\ref{eq:4Gab})-(\ref{eq:4Gabs}). 

Since 
$$\widehat{g}^{mn}=g^{mn}$$
i.e. there has been no field redefinition involving the four dimensional part of the metric, it is obvious that 
$$\frac{\de S_{\mt{grav}}}{\de g^{mn}}=  \widehat{G}_{mn}$$
\be
\Ra \frac{\de S_{\mt{grav}}}{\de g^{mn}}=0 \Longleftrightarrow \widehat{G}_{ab}=0
\ee
Thus we are left to show that 
$$\{\frac{\de S_{\mt{grav}}}{\de \Psi}=0,\frac{\de S_{\mt{grav}}}{\de \Te}=0\}\equiv \{\widehat{G}_{\as\bs}=0\}$$
A straight forward computation yields the field equations
$$\frac{\de S_{\mt{grav}}}{\de \Psi}=V_Ge^{-1}\Psi^{\Ds-1}\Te^{\Dd}\left\{\Ds R-2\Ds(\Ds-1)\Psi^{-1}\Box\Psi-2\Dd(\Ds-1)\Te^{-1}\Box\Te \right.$$
$$-\Ds(\Ds-1)(\Ds-2)\Psi^{-2}(\p\Psi)^2 
-\Dd(\Dd-1)(\Ds-2)\Te^{-2}(\p\Te)^2-2\Dd\Ds(\Ds-1)\Psi^{-1}\Te^{-1}\p_{a}\Psi \p^{a}\Te$$
\be
\left. +\4(\Ds-2)\Psi^{-2}\left[2\Dz+\Dd k\Te^{-2}-\2\Dz\Te^{2} \right]\right\} =0
\label{eq:4psi}
\ee
and 
$$
\frac{\de S_{\mt{grav}}}{\de \Te}=V_Ge^{-1}\Psi^{\Ds}\Te^{\Dd-1}\left\{\Dd R-2\Dd(\Ds-1)\Psi^{-1}\Box\Psi-2\Dd(\Dd-1)\Te^{-1}\Box\Te \right. $$
$$-\Dd(\Ds-1)(\Ds-2)\Psi^{-2}(\p\Psi)^2 
-\Dd(\Dd-2)(\Dd-1)\Te^{-2}(\p\Te)^2-2\Dd(\Dd-1)\Ds\Psi^{-1}\Te^{-1}\p_{a}\Psi \p^{a}\Te$$
\be
\left.+\Psi^{-2}\4\left[(2\Dz\Dd+\Dd k(\Dd-2)\Te^{-2}-\2\Dz(\Dd+2)\Te^{2}\} \right]\right\} =0
\label{eq:4theta}
\ee
Subtracting (\ref{eq:4theta}) from (\ref{eq:4psi}) gives us
$$\Dz\left[R-2(\Ds-1)\Psi^{-1}\Box\Psi-2\Dd\Te^{-1}\Box\Te-(\Ds-1)(\Ds-2)\Psi^{-2}(\p\Psi)^2 \right.$$
$$-\Dd(\Dd-1)\Te^{-2}(\p\Te)^2-2\Dd\Ds\Psi^{-1}\Te^{-1}\p_{a}\Psi \p^{a}\Te$$
$$
\left.-\2\Psi^{-2}(2-\Dz)+\4\Dd k\Psi^{-2}\Te^{-2}-\8(\Dz-4)\Psi^{-2}\Te^{2}\right]=0
$$
$$\Ra \widehat{G}_{\az\bz}=0$$
Also, by inspection 
$$(\ref{eq:4theta})\Ra \widehat{G}_{\ad\bd}= 0$$
We have thus succeeded in showing that the action (\ref{eq:4e-action}) is indeed consistent. It is easy to see that an addition of the cosmological term (\ref{eq:4c-action}) does not change the consistency of the truncation. 
\vspace{5mm}\\
\pagebreak\\
{\bf The Quantum Mechanical Action:} In the previous subsection we obtained the dimensionally reduced field theoretic action for our model. Since, we are not interested in the spatial fluctuations (which corresponds to particles) but rather the evolution of the background, we now assume the fields to only depend on time. In other words we use the fields as collective coordinates characterizing the observed and the internal space-time. For the internal space we already have 
\be
\psi(x)\ra S(t) \mbox{ and } \Te(x)\ra T(t)
\ee
 characterizing the size and the shape of the internal space respectively. For the external space we draw upon the standard  cosmological picture of an expanding universe:
\be
\mt{d}s^2=-e^{2W(t)}\mt{d}t^2+e^{2A(t)}\stackrel{\sqcup}{\dt}s^2
\label{eq:4replacement}
\ee
$A(t)$ is the usual cosmological radius of our universe while $W(t)$ corresponds to a gauge freedom which will be useful for later computations. We will also assume that the spatial metric $\st{\sqcup}{\mt{d}}s^2$ is flat, which recent observational data seem to suggest, and use the symbol $\sqcup$ to denote quantities corresponding to the space part of the observed space-time. Symbolically the full metric then looks like
\be
\widehat{\dt} s^2=-e^{2W(t)}\mt{d}t^2+e^{2A(t)} \stackrel{\sqcup}{\dt}s^2+e^{2S(t)}( \stc{\dt}s^2+e^{2T(t)} \sts{\dt}s^2)
\label{eq:4q-metric}
\ee

To obtain a quantum mechanical action from (\ref{eq:4e-action}) we basically need to calculate $R$ for the metric (\ref{eq:4q-metric}). Again, it is useful to cast the problem in terms of the vielbein. We define
\begin{equation}
e_m{}^a= \left( \begin{array}{cc}
e^{W(t)} & 0 \\
0  & e^{A(t)}\de_{\mmu}{}^{\au} 
\end{array} \right)
\end{equation}
We can now apply the same formalism as we used to calculate $\widehat{R}$. Alternatively, we can use conformal transformation by a scale factor $exp(A)$ to obtain $R$ from $R'=0$ for the trivial vielbein
$$
e'_m{}^a= \left( \begin{array}{cc}
e^{W(t)-A(t)} & 0 \\
0  & \de_{\mmu}{}^{\au} 
\end{array} \right)
$$
Anyway, one obtains 
\be
R=De^{-2W}[2\ddot{A} -2 \dot{A}\dot{W}+(D+1)\dot{A}^2]
\ee

Substituting $R$ and making the replacement (\ref{eq:4replacement})  we have the full quantum mechanical  action for the collective coordinates $W(t),A(t),S(t)$ and $T(t)$ from the effective gravitational action (\ref{eq:4e-action}).
\be
S_{\mt{qm,g}}=\int \mathtt{d} t\ e^{\Ds S+\Dd T+DA-W}[-K+V]
\label{eq:4q-action}
\ee
with
$$
K=-2D\ddot{A}-D(D+1)\dot{A}^2+\Ds(\Ds-1)\dot{S}^2+\Dd(\Dd-1)\dot{T}^2$$
\be
+2[\Dd(\Ds-1)\dot{T}\dot{S}+D\dot{A}\dot{W}]
\label{eq:4kinetic}
\ee
and
\be
V=\4e^{2(W-S)}\left[2\Dz+\Dd ke^{-2T}-\2\Dz e^{2T}\right]
\ee
It is also simple to include the cosmological term (\ref{eq:4c-action}) in the quantum mechanical  action
\be
S_{\mt{qm,c}}=-2\La\int \mathtt{d} t\ e^{\Ds S+\Dd T+DA+W}
\ee
so that the total action becomes
\be
S_{\mt{qm,eff}}=S_{qm,g}+S_{qm,c}
\ee

At this point it is tempting to identify 
\be
V_{\mt{ eff }}= \4e^{(\Ds-2)S+\Dd T+DA)}\left[2\Dz+\Dd ke^{-2T}-\2\Dz e^{2T}\right]-2\La e^{\Ds S+\Dd T+DA}
\ee
as the effective potential in the usual gauge $W=0$. However, a quick analysis will reveal, that for example 
$$
\frac{\p V_{\mt{ eff }}}{\p S}=\frac{\p V_{\mt{ eff }}}{\p T}=0
$$
does not give us the correct parameter values for the Kaluza-Klein vacuum, the correct values of $S$ and $T$ being those for which the internal manifold is Einstein. The essential reason for this anomaly is also simple; $ S_{\mt{ qm,eff }}$ is a non-linear sigma model with non-trivial prefactors in the kinetic terms (\ref{eq:4kinetic}). To extract more information about the dynamics we need to perform a more careful analysis.
\setcounter{equation}{0}
\section{ {\bf FIELD EQUATIONS}}
{\bf The Gauge Choice:} Inspection of the action (\ref{eq:4q-action}) tells us that it greatly simplifies if we choose the gauge
\be
W=DA+\Ds S+\Dd T
\ee
We no longer have a non-linear sigma model, but rather a sum of ordinary kinetic terms. One can always transform back the result to the more familiar gauge $W=0$. A similar gauge  was recently used \cite{kaya} in the context of brane gas cosmology at the level of field equations.  In this ``linear gauge'' the effective action becomes
\be
S_{\mt{qm}}=\int \mathtt{d} t\ [K_{\mt{qm}}-V_{\mt{qm}}]
\label{eq:4qe-action}
\ee
with
$$
K_{\mt{qm}}=D(D-1)\dot{A}^2+\Ds(\Ds-1)\dot{S}^2+\Dd(\Dd-1)\dot{T}^2$$
\be
+2[D\Ds\dot{A}\dot{S}+D\Dd\dot{A}\dot{T}+\Dd(\Ds-1)\dot{T}\dot{S}]
\ee
and
\be
V_{\mt{qm}}=\4e^{2(DA+(\Ds-1)S)}\left[2\Dz e^{2\Dd T}+\Dd ke^{2(\Dd-1)T}-\2\Dz e^{2(\Dd+1)T}\right]-2\La e^{2(\Ds S+\Dd T+DA)}
\label{eq:4q-potential}
\ee

In obtaining the effective quantum mechanical action we have ignored the total derivative terms and dropped some prefactors. By plotting the  effective quantum mechanical potential $V_{\mt{qm}}$ one can perhaps get a good qualitative idea about the dynamics  but we are primarily interested in the behaviour of $T(t)$, the transition function, and there are at least two reasons why $V_{\mt{qm}}$ does not provide an accurate description of the same. Firstly, there are still mixed kinetic  terms which doesn't allow an easy interpretation of the potential. Secondly and more importantly, we are plagued with the usual problem of stability of the size of the internal manifold. It is well known that the internal manifold in the Kaluza-Klein picture tends to grow. There have been several attempts to stabilize the vacuum in the supergravity/string theory framework. In particular, brane gas cosmology \cite{brane,easson,kaya} has achieved reasonable success in explaining how the internal manifold can remain stable, while the external universe can expand. Clearly our model of pure gravity should only be viewed as a toy model and in a realistic theory one needs to not only include matter/radiation but also a mechanism to stabilize the size. Thus the potential $V_{\mt{qm}}$ is incomplete and since the dynamics of $S$ and $T$ are intricately connected, the reliability of the information that it can provide for $T$ is in question. 

To do an honest job one therefore needs to include the different contributions to the stress energy tensor coming from different forms of matter and energy. However, as we shall see there is still a way out of this impasse which lets us to at least get a qualitative idea as to how the squashing parameter will behave. The trick is to work at the level of field equations. 
\vspace{5mm}
\\ 
{\bf Field Equations:} The field equations that one gets by varying the action (\ref{eq:4qe-action}) reads
$$
\frac{\de S_{\mt{qm}}}{\de A}=0\ \Ra \ 2(D-1)\ddot{A} +2\Ds\ddot{S}+2\Dz\ddot{T}$$
$$+e^{2(DA+(\Ds-1)S+\Dd T)}\left[\Dz +\2\Dd ke^{-2T}-\4\Dz e^{2T}\right]-4\La e^{2(DA+\Ds S+\Dd T)}=0$$
$$
\frac{\de S_{\mt{qm}}}{\de S}=0\ \Ra \ 2D\Ds\ddot{A} +2\Ds(\Ds-1)\ddot{S}+2\Dz(\Ds-1)\ddot{T}$$
$$+\2 (\Ds-1)e^{2(DA+(\Ds-1)S+\Dd T)}\left[2\Dz +\Dd ke^{-2T}-\2\Dz e^{2T}\right]-4\La\Ds e^{2(DA+\Ds S+\Dd T)}=0$$
$$
\frac{\de S_{\mt{qm}}}{\de T}=0\ \Ra \ 2D\Dd\ddot{A} +2(\Ds-1)\Dd\ddot{S}+2\Dz(\Dz-1)\ddot{T}$$
$$+\4 e^{2(DA+(\Ds-1)S+\Dd T)}\left[4\Dz\Dd +2\Dd(\Dd-1) ke^{-2T}-\Dz(\Dd+1) e^{2T}\right]-4\La\Dd e^{2(DA+\Ds S+\Dd T)}=0$$
A simple rearrangement of the equations gives us
\be
\ddot{A}-\frac{4\La}{\Dh-2} e^{2(DA+\Ds S+\Dd T)}=0
\label{eq:4A}
\ee
\be
\ddot{S}+e^{2DA}\left\{\2 e^{2(\Ds-1)S}\left[ e^{2\Dd T}-\2 e^{2(\Dd+1)T}\right]-\frac{4\La}{\Dh-2} e^{2\Ds S}e^{2\Dd T}\right\}=0
\label{eq:4S}
\ee
and
\be
\ddot{T}+\2 e^{2(DA+(\Ds-1)S)}\left[\2 ke^{2(\Dd-1)T}- e^{2\Dd T}+\4 (1+\textstyle{\Ds\over\Dd}) e^{2(\Dd+1)T}\right]=0
\label{eq:4T}
\ee
One can immediately find the vacuum solutions, i.e. when the internal manifold is frozen. For constant $S$ and $T$, $\ddot{S}=\ddot{T}=0$ and from (\ref{eq:4S}) and (\ref{eq:4T}) we have
\be
 ke^{-2T}- 2+\2 (1+\frac{\Ds}{\Dd}) e^{2T}=0
\label{eq:4soln}
\ee
and
\be
\2 \left[ 1-\2 e^{2T}\right]-\frac{4\La}{\Dh-2} e^{2 S}=0
\ee
Substituting $k$ in (\ref{eq:4soln}) we have
$$
\left(1-\frac{\Dz}{2\Dd}\right)e^{-2T}- 2+ \left(1+\frac{\Dz}{2\Dd}\right) e^{2T}=0
$$
This has two solutions
\be
e^{2T}=1,\ \frac{2\Dd-\Dz}{2\Dd+\Dz}
\ee
and correspondingly
\be
e^{2S}=\frac{\Dh-2}{16\La},\ \left(\frac{\Dh-2}{16\La}\right)\left(\frac{2\Dd+3\Dz}{2\Dd+\Dz}\right)
\ee
Indeed these are the right vacuum solutions for the full higher dimensional Einstein's equations. The first one corresponds to the symmetric case, while the second one to the squashed case. 

Are these vacua stable? Is a vacuum to vacuum transition possible? Let us try to investigate these issues in the next section.
\setcounter{equation}{0}
\section{{\bf QUINTESSENTIAL PHASE TRANSITION }}
{\bf Effective Potentials:} In the previous section we have obtained equations of motion for the collective coordinates in a form reminiscent of Newton's second law (\ref{eq:4A})-(\ref{eq:4T}), except of course the force being coupled. However, the expression for force in (\ref{eq:4T}) is especially nice, because it has factorized enabling us to have a qualitative picture of the dynamics of $T$. Treating $S$ and $A$ constant for the moment, we can define
\be
V_{\mt{eff}}(T)=c_1\left[\frac{1}{4(\Dd-1)} \left(1-\frac{\Dz}{2\Dd}\right)e^{2(\Dd-1)T}- \frac{1}{2\Dd}e^{2\Dd T}+\frac{1}{4(\Dd+1)} \left(1+\frac{\Dz}{2\Dd}\right) e^{2(\Dd+1)T}\right]
\ee
such that 
\be
\ddot{T}+\frac{\p V_{\mt{eff}}}{\p T}=0
\ee
Here we have  treated
$$
c_1=\2 e^{2(DA+(\Ds-1)S)}$$
as a constant which is obviously untrue. However, $V(T)$ should be able to explain the dynamics approximately\footnote{Although evident, it should perhaps be emphasized that the potential $V(T)$ cannot be directly identified as the scalar field potential that is usually used to analyse inflation or quintessence in the literature and hence one cannot straightforwardly apply previously known results.}. For example, one finds the correct vacuum values just by looking at the minima/maxima of the potential $V(T)$.

By looking at $V(T)$ we find that there are two possible cases:\\
(a) $2\Dd\ \leq \Dz$: There is only one vacuum solution\footnote{The squashed solution being negative and hence unphysical.} corresponding to the minima of the potential $V(T)$. A typical example would be when $G=SU(2)$ and $H=U(1)$.
\begin{center}
\scalebox{.7}{\includegraphics*[3in,1.6in][7in,5in]{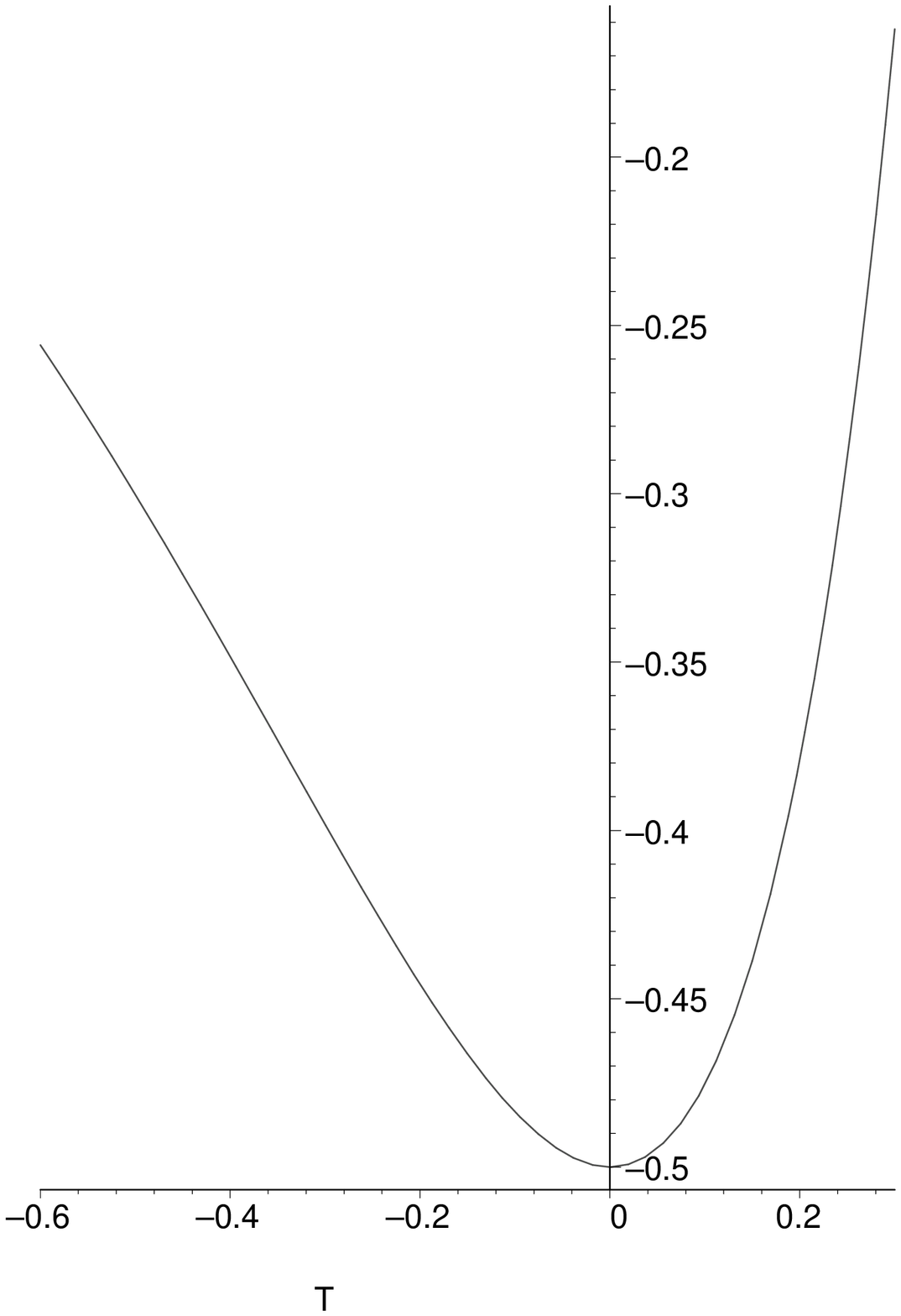}}\\
$\st{\longrightarrow}{T}$\\
figure 1: $SU(2)\longrightarrow U(1)$\\
\end{center}
 As is clear from the plot (fig. 1), no transition is possible and correspondingly the gauge group $SU(2)$ cannot be broken to $U(1)$. \\
(b) $2\Dd \ > \Dz$: In this case there are two vacuum solutions. Typical examples are  $G=SU(3)\ra H=SO(3)$ and $G=SO(4)\ra H=SO(3)$. 
\begin{center}
\scalebox{.4}{\includegraphics*[1in,1.6in][9in,10in]{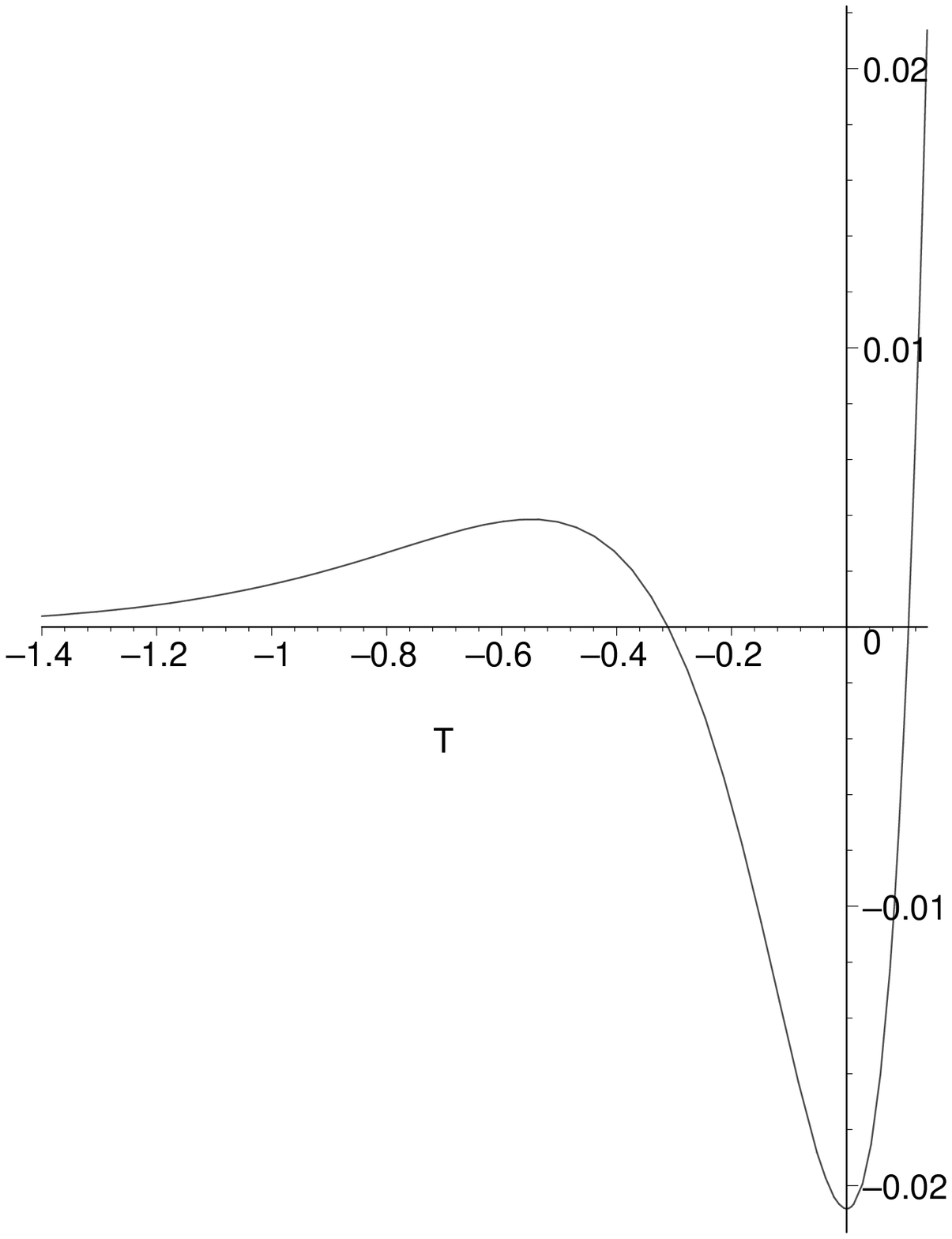}}\\
figure 2: $SO(4)\longrightarrow SO(3)$\\
\end{center}
As the plot shows (fig. 2), while the symmetric vacuum corresponds to a minima in $V(T)$, the squashed vacuum corresponds to a maxima and is therefore unstable. Thus the symmetric-vacuum to squashed-vacuum transition as suggested in \cite{t} is untenable\footnote{Things may be different if either $G$ or $H$ are not simple, or if the coset decomposition is not symmetric etc.}  unless the squashed vacuum is somehow stabilized (perhaps by supersymmetry \cite{stability}). However, as is clear from the plot, there can still be a phase transition! Moreover, such a cosmological transition could be very interesting because it may be able to explain inflation and/or quintessence.
\vspace{5mm}
\\ 
{\bf Inflation:} Consider first the inflationary scenario. It is natural to assume that the universe started out in a symmetric  phase at the minima of $V(T)$. $T$ however could be displaced because of classical or quantum fluctuations/excitations and start to roll over or tunnel through the potential barrier. During this phase of rolling over we will have inflation. At the squashed vacuum one can indeed solve Einstein's equations and in the  $W=0$ gauge   one finds
\be
e^{2A(t)}=e^{\Gamma_{\mathtt{I}} t} \mx{ with exponent } \Gamma_{\mathtt{I}}=\sqrt{\frac{8\La}{D(\Dh-2)}}
\ee
an inflationary growth. This solution is none other than the deSitter vacuum $dS_{D+1}\otimes G_{\mt{sq}}$, which in our dynamic universe model is just a phase. Exponential potentials have indeed been studied in the context of inflation with considerable success \cite{maeda} which augers well for our model. 

Once the potential has dropped sufficiently, inflation stops. By now most of $T$'s potential energy would have been converted into kinetic energy which is eventually converted to radiation/matter (see for example \cite{spokoiny, peebles}) and the universe will presumably enter into  a radiation-matter dominated phase. At this time $A$ should scale as the radiation and matter dominated era which is previously known to happen for exponential potentials \cite{liddle,copeland}. 
\vspace{5mm}
\\ 
{\bf quintessence:} In \cite{spokoiny}, Spokoiny already considered the conditions when the inflaton potential energy can again dominate the expansion of the universe. This found a logical culmination in the idea of quintessential inflation \cite{peebles} where the inflaton field plays the dual role of a quintessence-field. This is possible in the later stages of the evolution of the inflaton if it keeps rolling towards infinity rather than settling down in another minima. The potential energy, although appreciably smaller than the initial inflation energy, starts to dominate again as the matter and radiation density drops. It acts as an  effective cosmological constant (quintessence) thereby causing the universe to accelerate which recent observational data seem to suggest \cite{exp}. Such a mechanism can naturally explain the smallness of the effective cosmological constant and hence have received much attention lately. Several authors have studied  exponentially falling potentials \cite{liddle} in this context and in particular, in \cite{copeland} it was shown how a linear combination of exponential potentials may just be perfect for providing a satisfactory explanation of most of the cosmological data that is available. In our model $V(T)$ is indeed a linear combination of exponential potentials and as $T$ rolls to $-\infty$, quintessence becomes a viable prospect. The symmetry breaking transition is therefore not an ordinary vacuum to vacuum transition but rather a transition from a vacuum to quintessence, and hence I call it a quintessential transition!

Let us see whether our toy model of pure Kaluza-Klein theory can provide a ``quintessential solution'' of an accelerating universe. In the later phase of evolution for sufficiently small $T$, the smallest exponent in the potential $V(T)$ dominates, $T$ essentially rolling down  $e^{2(\Dd-1)T}$. Thus we can ignore all the other terms in the effective action (\ref{eq:4qe-action}-\ref{eq:4q-potential}). This effectively conceals the curvature of $G/H$ and we are left with the product space $M_{D+1}\otimes (G/H)_{\Dz}\otimes H$, where $H$ acquires an internal curvature $(1/4)k$ while $G/H$ becomes flat. It is easy to find a deSitter type solution:
\be
S(t)=A(t) \mx{ and } T(t)=T_0-A(t)
\label{eq:4Q-soln}
\ee
 In the $W=0$ gauge 
\be
e^{2A(t)}=e^{\Gamma_{\mathtt{q}} t} \mx{ with exponent } \Gamma_{\mathtt{q}}=\sqrt{\frac{8\La}{(D+\Dz)(\Dh-2)}}
\label{eq:4exponent}
\ee
We notice that the quintessence exponent is smaller than the inflation exponent as it should be. 

   One can be skeptical of the exponential growth of $S$ in the quintessence solution (\ref{eq:4Q-soln},\ref{eq:4exponent}), but this has a clear explanation. As discussed earlier, this is the usual problem of Kaluza-Klein cosmology as can also be seen from the effective potential of $S$ that one can define from (\ref{eq:4S}). 
\be
V_{\mt{eff}}(S)= c_2\left[\frac{1}{4(\Ds-1)} e^{2(\Ds-1)S}-\frac{2\La}{\Ds(\Dh-2)} e^{2\Ds S}\right]
\ee
where we have chosen $A$ and $T\ra (-\infty)$ to be constants. A typical plot looks like
\begin{center}
\scalebox{.5}{\includegraphics*[1in,1.6in][7in,10in]{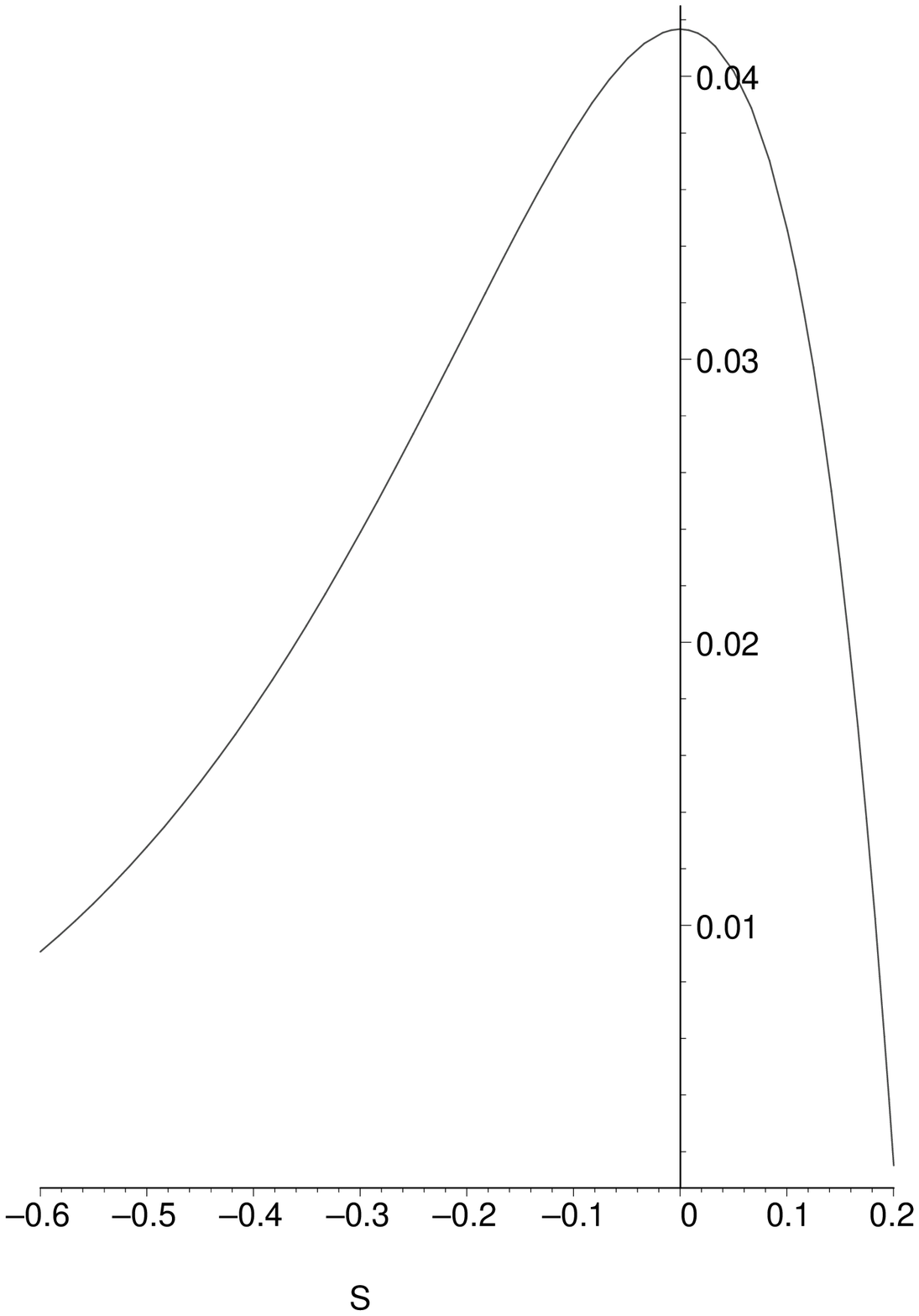}}\\
$\st{\longrightarrow}{S}$\\
For $\La=1$ and $T\ra \infty $\\
\end{center}
Clearly, the equilibrium point is a maximum and hence unstable\footnote{This also shows how $V(T)$ and $V(S)$ can provide a good picture of the dynamics involved}. For Kaluza-Klein cosmology to be consistent, the size of the internal manifold has to stabilize and recent advances in brane gas cosmology is promising in this regard. Thus, for a comprehensive study of this model we have to also include a stabilising mechanism for $S$. All these ``inclusions'' will intricately effect the $T$-dynamics and in fact it may so happen that instead of $T$, a linear combination of $S$ and $T$, or may be both be involved in inflation and quintessence. 

\section{{\bf SUMMARY AND FUTURE RESEARCH}} 

In \cite{t}, I had tried to explain how a dynamical internal manifold can break gauge symmetry partially. In particular, a transition of the internal manifold from a symmetric vacuum to a squashed, and hence less symmetric vacuum, was suggested. In this paper I have tried to analyze in more detail when and how such a transition can occur. For simplicity, we focused on the case when the internal manifold is a simple Lie group $G$ and we are interested in breaking the isometry group from $G_R \ra H_R$. Further, we assumed $H$ to be simple and the coset space $G/H$ to be reductive and symmetric. We studied the dynamics using collective coordinates characterizing the size (radius) and the shape (squashing parameter)  of the internal manifold. In particular, we derived an effective potential for the squashing parameter which gives a nice qualitative picture of the dynamics. We observed that there are two possible cases depending upon the dimensions $\Dz$ of $G/H$ and $\Dd$ of $H$: When $\Dz\ \geq \2\Dd$, we found that no transition can occur because there is only a single minima. This explains why we can't have a transition from $SU(2)$ to $U(1)$ for example. However when $\Dz\ < \2\Dd$, a nontrivial transition can indeed occur. In this case there are two equilibrium points in the potential, a minima (symmetric vacuum) and a maxima (squashed vacuum). Thus although a symmetric-vacuum to squashed-vacuum transition is unfeasible, the squashed parameter can role over or tunnel through the potential barrier. It then keeps on evolving toward $-\infty$, something similar to what a quintessence field does in the later stages of its evolution. Hence I believe it is appropriate to call this phenomena  a quintessential transition. Since, the squashing field goes further and further away from the symmetric vacuum it cannot tunnel or roll back toward it. 

Such a transition can have profound cosmological consequences. It may be able to explain inflation and/or quintessence. The rolling over phase of the squashing field is accompanied by inflation or an exponential growth in the radius of the external universe.  Following inflation the potential energy of the squashing field will presumably become subdominant to the matter-radiation contributions, but it can resurface again at a later stage. This is essentially the scenario of quintessential inflation, where the inflaton at a later stage of evolution can account for a small effective cosmological constant. Exponential potentials have been previously studied in the context of both inflation \cite{maeda} and quintessence \cite{liddle,copeland}  and we do indeed obtain exponential potentials in our model. We saw in a toy example, that as the squashing field is slowly rolling down the potential hill, the internal manifold getting more and more squashed, it is indeed accompanied by an exponential growth in the cosmological radius of the observed four dimensional universe. This quintessence exponent is smaller than the inflationary exponent as one expects.

There are still a lot of unanswered questions. In order for our model to be feasible   it has to  satisfy several constraints coming from observational cosmology. We have to  include matter-radiation effects in future calculations. Further, consistent Kaluza-Klein cosmology would almost certainly require a mechanism to stabilize the radion or size of the internal manifold, like presence of brane gas, and these effects have also to be factored in. Last but not the least, a nice feature about this model is that it combines a fundamental phenomena in particle physics, that of symmetry breaking in gauge theories, with cosmological effects. This theory naturally predicts the mass of the broken gauge bosons which only depends on group theoretic parameters, and the current size and shape of the internal manifold.  Thus for this model to be truely successful it should also be able to explain the values of the vector bosons coming from the particle physics estimates.

Finally, several variations of the same idea can and should be studied for realistic phenomenological applications. Firstly, one can generalize the model from pure gravity to supergravity. The extra fields (dilaton and/or various form fields) may contain   scalars that are relevant to the squashing dynamics. Secondly, it should be possible to generalise the internal manifolds and metrics that I considered in this paper which can change some of the parameters and even the nature of the effective potentials. For example tt should  not be too difficult to generalise this mechanism to more realistic internal manifolds like the coset spaces (and in particular those which can give Standard Model like gauge groups), and at least in principle to some of the more interesting inhomogeneous spaces. 
\vspace{5mm}\\
{\large {\bf Acknowledgments:}} I would like to thank Prashanth Jaikumar for some  useful discussions and suggestions.

\end{document}